\documentclass{caosp309}
\usepackage{graphicx}
\usepackage{natbib}
\usepackage{amsmath,amssymb,enumitem}
\usepackage[hyphenbreaks]{breakurl}
\articleNo{1}
\pubyear{2023}
\volume{1}
\volnumber{1}
\firstpage{1}
\received{2023}
\accepted{2023}
\def\BibTeX{{\rm B\kern-.05em{\sc i\kern-.025em b}\kern-.08em
             T\kern-.1667em\lower.7ex\hbox{E}\kern-.125emX}}

\begin{document}

\htitle{Study of spectral index of giant radio galaxy from Leahy's Atlas: DA 240}
\hauthor{V. Borka Jovanovi\'{c}, D. Borka and P. Jovanovi\'{c}}

\title{Study of spectral index of giant radio galaxy from Leahy's Atlas: DA 240}

\author{V. Borka Jovanovi\'{c}\inst{1}\orcid{0000-0001-6764-1927}
\and D. Borka\inst{1}\orcid{0000-0001-9196-4515}
\and P. Jovanovi\'{c}\inst{2}\orcid{0000-0003-4259-0101}}

\institute{Department of Theoretical Physics and Condensed Matter Physics (020), Vin\v{c}a Institute of Nuclear Sciences - National Institute of the Republic of Serbia, University of Belgrade, P.O. Box 522, 11001 Belgrade, Serbia, \email{vborka@vinca.rs}, \email{dusborka@vinca.rs}
\and Astronomical Observatory, Volgina 7, P.O. Box 74, 11060 Belgrade, Serbia, \email{pjovanovic@aob.rs}}

\date{March 8, 2003}
\maketitle

\begin{abstract}
Here we investigate the giant radio galaxy DA 240, which is a FR II source. Specifically, we investigate its flux density, as well as the spectral index distribution. For that purpose, we used publicly available data for the source: Leahy's atlas of double radio-sources and NASA/IPAC Extragalactic Database (NED). We used observations at 326 MHz (92 cm) and at 608 MHz (49 cm) and obtained spectral index distributions between 326 and 608 MHz. For the first time we give spectral index map for these frequencies. We found that the synchrotron radiation is the dominant radiation mechanism over most of the area of DA 240, and also investigated the mechanism of radiation at some characteristic points, namely its core and the hotspots. The results of this study will be helpful for understanding the evolutionary process of the DA 240 radio source.
\keywords{galaxies: active -- galaxies: jets -- galaxies: nuclei -- radio continuum: galaxies -- galaxies: individual: DA 240}
\end{abstract}

\section{Introducion}

Double Radio sources Associated with Galactic Nuclei (DRAGNs) are clouds of radio-emitting plasma which have been shot out of active galactic nuclei (AGN) via narrow jets. More precisely, a DRAGN would be a radio source containing at least one of the following types of extended, synchrotron-emitting structures: jet, lobe, and hotspot complex \citep{leah93}. 

AGNs are emitting the most radiation from galaxies, and in case of radio galaxies, a lot of their radiation is emitted at radio wavelengths. Giant radio galaxies (GRGs) belong to a unique class of objects with very large radio structures. Originally, they were defined to be the radio galaxies with projected linear sizes greater than 1 Mpc \citep{will74}. This limit applied to a spatially flat ($\Omega_\kappa = 0$) Friedmann cosmological model with the Hubble constant H$_0$ = 75 km s$^{-1}$ Mpc$^{-1}$, deceleration parameter $q_0 = 0.5$ and zero cosmological constant ($\Omega_\Lambda = 0$). Nowadays, the GRG size limit is equivalent to 700 kpc \citep{tang21} in a $\Lambda$CDM cosmology with the parameters by Planck Collaboration from 2016, i.e. in the flat cosmological model with H$_0$ = 67.8 km s$^{-1}$ Mpc$^{-1}$ and $\Omega_m = 0.308$ \citep{plan16}. Due to this comparatively large angular extent of the GRG population, astronomers can observe their fine structures with detailed imaging.

DRAGN DA 240 was among the first GRGs to be recognized as such (more precise, 3C 236 and DA 240 are the first two discovered GRGs, both identified as Fanaroff-Riley II types). A study of its environment can be found in \citet{peng04}. It consists of two radio clouds about 40$^\prime$ long, and a comparatively weak central core \citep{arty88}. This giant radio source, with linear size spanning over 1.3 Mpc, is placed at a distance of 215 Mpc. There are also other researchers who investigated radio source DA 240 \citep{mack97,chen11a,chen11b,chen18a,chen18b,peng15,mile19}.

\section{Data and method}

Regarding Earth-space (and also space-Earth) communication ran-ges, there is a wide spectral window in radio band in which Earth's ionosphere does not reflect extraterrestrial radio waves and where atmosphere is transparent for them. This range of radio frequencies, observable from Earth, spans from $\sim$10 MHz (30 m) to $\sim$1 THz (0.3 mm). Particularly, for DA 240 there are freely available ground-based observational data at the following frequencies: 326, 608, 2695, 4750 and 10550 MHz. Among these, for this paper we choose observations only at 326 and 608 MHz because of their completeness, and due to good visibility of this source over the whole its area, as well. This is not the case for other three frequencies, where only the regions around hotspots and jets are clearly visible, while the radio lobes are pretty faint at these frequencies.

For our calculations, we used Flexible Image Transport System (FITS) data files containing the flux densities in Jy (1 Jy = $10^{-26}$ W m$^{-2}$ Hz$^{-1}$) of a chosen radio source. We described the structure of FITS format, as well as what is useful for our investigation, in our previous paper \citet{bork23}. Besides its flexibility and storage efficiency, here we want to point out the long term archiving i.e. all versions of the FITS format are backwards-compatible, so one can compare data from more observations (details at \url{https://heasarc.gsfc.nasa.gov/docs/heasarc/fits_overview.html}).

The observed data are available in \texttt{An Atlas of DRAGNs} i.e. the ''3CRR'' sample of \citet{lain83}. Besides FITS files, this sample gives the readers very useful information from the literature on the DRAGNs, with tables and references, too. There are the Introductory Pages, Description Pages (with full details) and the Listings of individual DRAGNs. Also, we used astronomical database compiled by NASA and IPAC, i.e. NED database - a comprehensive database of multiwavelength data for extragalactic objects, with the information and bibliographic references regarding these objects.

Hence, the easily searchable and accessible data are provided at:

\sloppy
\begin{itemize}
\item J. P. Leahy, A. H. Bridle, R. G. Strom, An Atlas of DRAGNs (2013): \url{http://www.jb.man.ac.uk/atlas/} \citep{leah13},
\item NASA/IPAC Extragalactic Database: \url{http://ned.ipac.caltech.edu/} \citep{mazz02}.
\end{itemize}
\fussy

The observations of DA 240 at 326 and 608 MHz were carried out using ''The Westerbork Synthesis Radio Telescope'' (WSRT) radio telescope. It is located in Netherlands, previously ran by Netherlands Foundation for Research in Astronomy, underwent more upgrades and a major (phased array upgrade of the WSRT) was completed in 2019, resulting in the WSRT-Apertif system. Nowadays, WSRT-Apertif telescope is operated as a survey instrument by ASTRON -- the Netherlands Institute for Radio Astronomy.

It is useful here to notice that we used the calibrated data, i.e. the data and images after they are processed using analysis techniques and programs like \texttt{CLEAN} algorithm (see \cite{stro81} and references therein). Particularly, regarding the resolutions of the processed data used here, their values are: 20$''$ at 326 MHz and 9.2$''$ at 608 MHz.

We used observations of DA 240 at two frequencies, 326 MHz \citep{will90} and 608 MHz \citep{will74}. The calculation method, which we have developed, was first published in \citet{bork07}, with the most detailed explanation given in \citet{bork12a}, and further elaborated in \citet{bork12b}, also. The area of the investigated radio source, as well as the flux densities, are determined in these three ways: I - flux density contours (isolines $S_\nu$); II - flux density 2D profiles, for constant declination; III - 3D profiles (by doing the procedure in all three ways, it can easily be checked whether the results and analysis are good).

As a telescope is most sensitive to extended emission at long wavelengths (i.e. smaller frequencies), the WSRT could provide detailed and sensitive maps of DA 240 at long wavelengths. So, it had a resolvong power sufficient to separate this large object from unrelated confusing sources. As noted in original discovery paper by \citet{will74}, a remarkable feature is the huge range of surface brightness over the intensity map, as well as the prominence of the eastern component. Also, it has strong linear polarization: an integrated percentage polarization at 49 cm of 8.3\%.

\clearpage
\section{The flux density distribution of giant radio galaxy DA 240}

The radio source DA 240 (cross-identifications: DA 240; CGCG 262-029; CGCG 0744.6+5556; MCG +09-13-057; 4C +56.16; 2MASX J07483682+5548591) is an example of Classical Double radio sources. Its projected linear size is about 1350 kpc.

\begin{figure*}[ht!]
\centering
\includegraphics[width=0.50\textwidth]{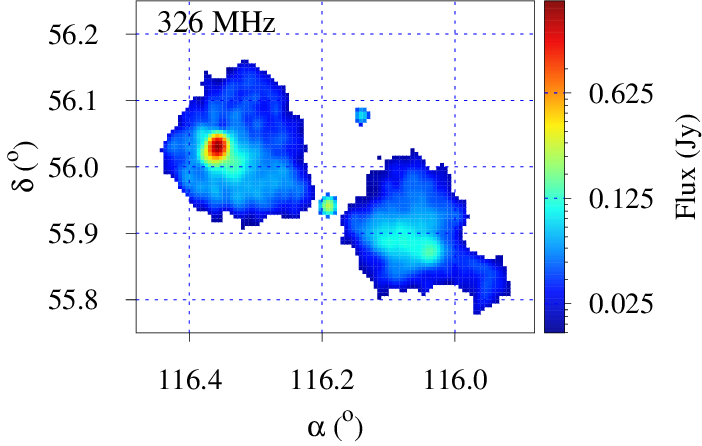}
\hfill
\includegraphics[width=0.48\textwidth]{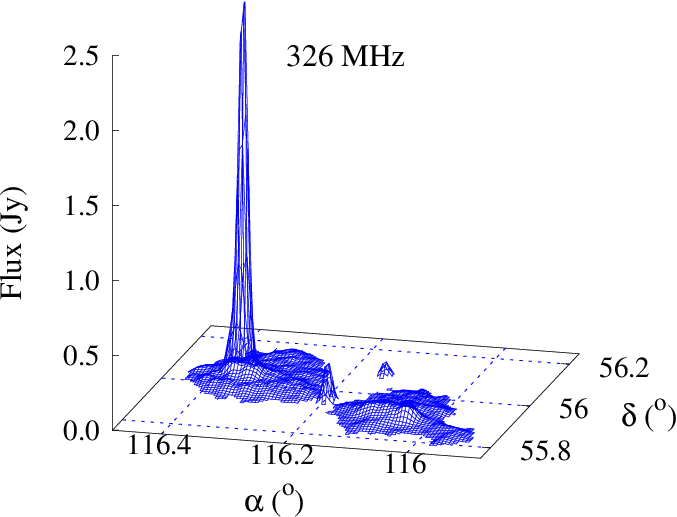}
\caption{2D plot (left) and 3D plot (right) of DA 240 flux density distribution (in Jy), presented in Equatorial coordinate system $(\alpha, \delta)$, at 326 MHz.}
\label{fig01}
\end{figure*}

\begin{figure*}[ht!]
\centering
\includegraphics[width=0.50\textwidth]{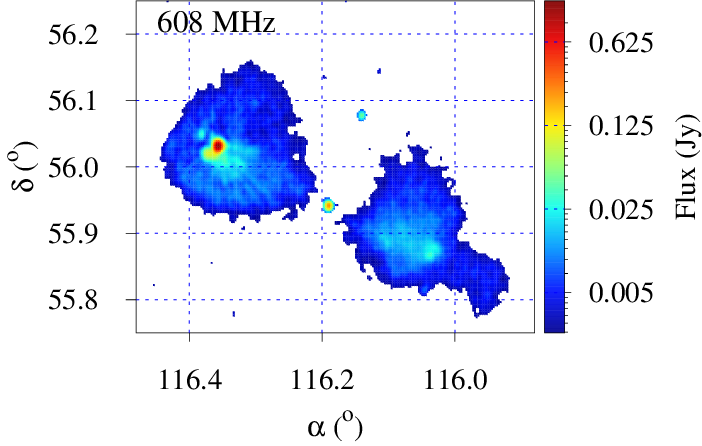}
\hfill
\includegraphics[width=0.48\textwidth]{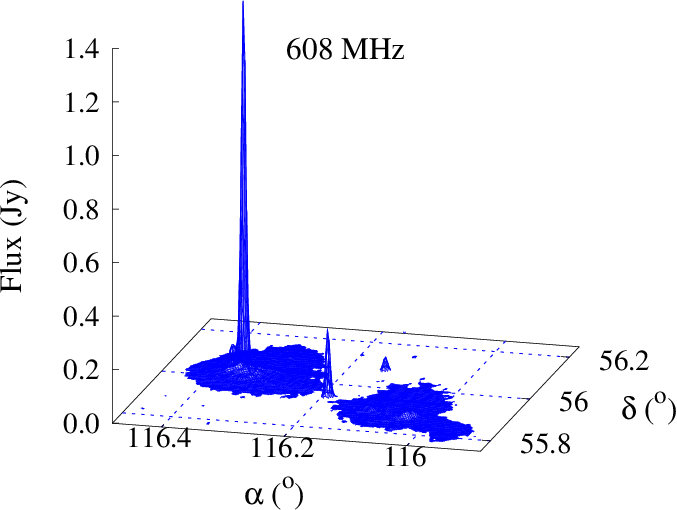}
\caption{The same as in Fig. \ref{fig01}, but for 608 MHz.}
\label{fig02}
\end{figure*}

From the observations of DA 240 at two frequencies, 326 MHz (92 cm) and 608 MHz (49 cm), we determined the contours which represent the lower boundaries of the source. We found that the minimal fluxes are the following: $S_{\nu,min}$ = 0.016 Jy at 326 MHz, and $S_{\nu,min}$ = 0.0023 Jy at 608 MHz. The areas of DA 240, with the flux density distributions (in Jy) over these areas, we presented by two-dimensional and three-dimensional plots in Figs. \ref{fig01} and \ref{fig02}. From these radio maps, the spherical radio lobes can be clearly seen, as well as their hotspots at the end of each beam.

\begin{figure*}[ht!]
\centering
\includegraphics[width=0.48\textwidth]{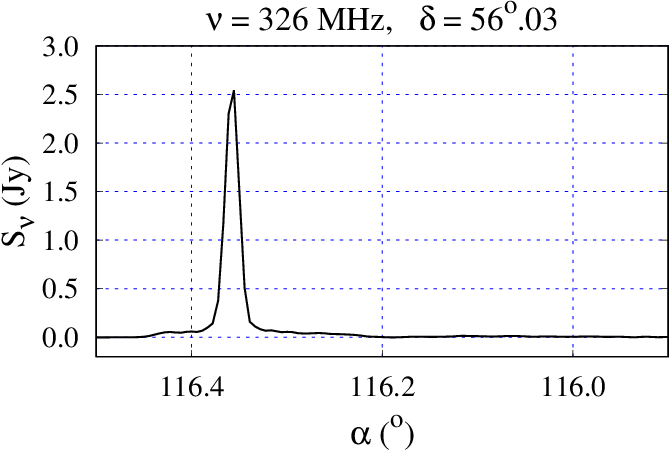}
\hfill
\includegraphics[width=0.48\textwidth]{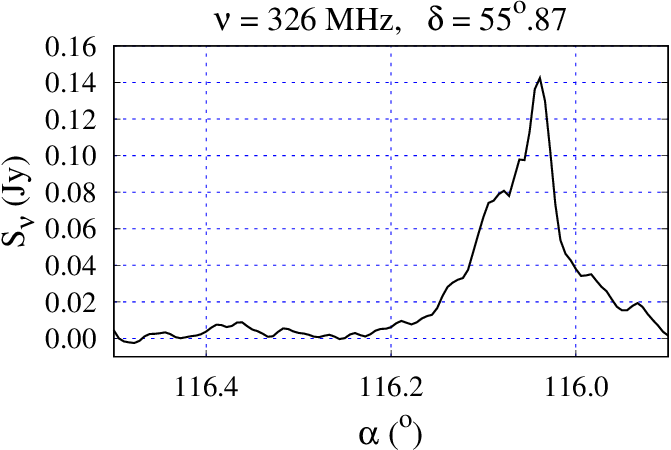}
\caption{The 326 MHz flux profiles for constant declinations $\delta = 56^\circ.03$ (left) and $\delta = 55^\circ.87$ (right), containing northern and southern hotspots.}
\label{fig03}
\end{figure*}

\begin{figure*}[ht!]
\centering
\includegraphics[width=0.48\textwidth]{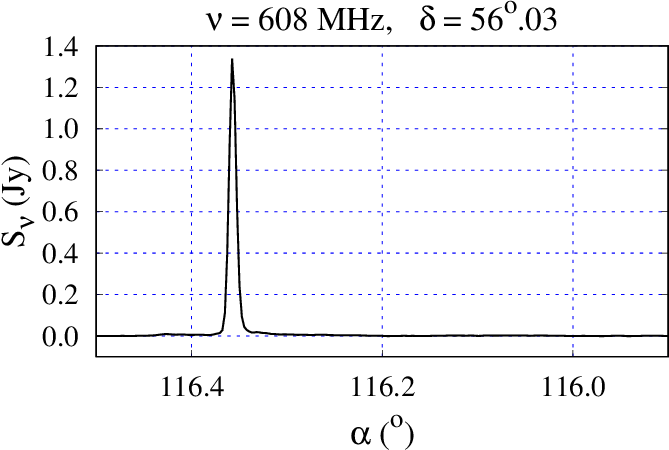}
\hfill
\includegraphics[width=0.48\textwidth]{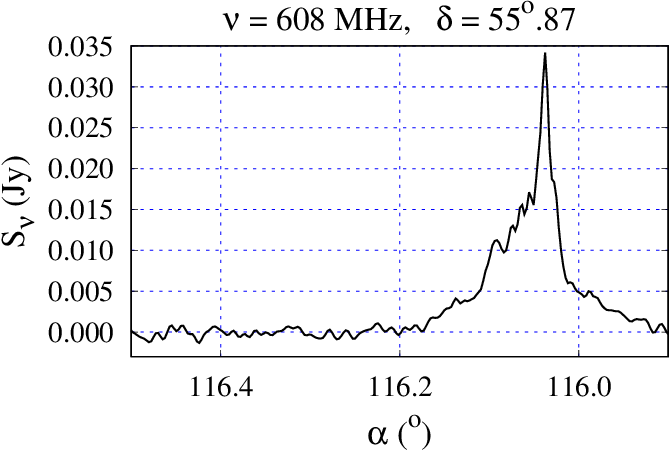}
\caption{The same as in Fig. \ref{fig03}, but for 608 MHz.}
\label{fig04}
\end{figure*}

We also give an examples of flux profiles for some constant declinations, at the both frequencies. In Fig. \ref{fig03} we give the profile for $\delta = 56^\circ.03$ containing north-eastern hotspot (left) and for $\delta = 55^\circ.87$ containing south-western hotspot (right), at 326 MHz; while in Fig. \ref{fig04} we give profiles for the same $\delta$ but at 608 MHz.

\section{Spectral index distribution between 326 and 608 MHz}

The amount of flux density $S_\nu$ as a function of frequency $\nu$ is given by the expression:

\begin{equation}
S_\nu \sim \nu^{-\alpha},
\label{equ01}
\end{equation}

\noindent where $\alpha$ is a constant, called the ''radio spectral index''.

The radio spectral index $\alpha$ can be obtained using the flux density at different frequencies and taking the negative slope of the relation (\ref{equ01}). Therefore, we calculate it by the following equation:

\begin{equation}
\alpha = - \dfrac{\log\left({\dfrac{S_{\nu_1}}{S_{\nu_2}}}\right)}{\log\left({\dfrac{\nu_1}{\nu_2}}\right)}.
\label{equ02}
\end{equation}

\begin{figure*}[ht!]
\centering
\includegraphics[width=0.85\textwidth]{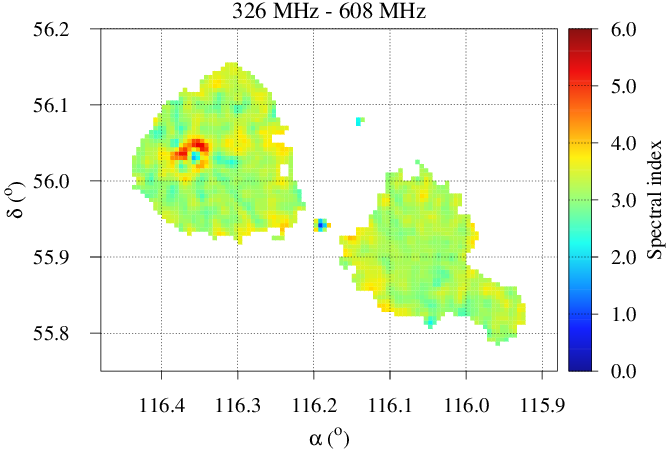}
\caption{Spectral indices between 326 and 608 MHz over the area of DA 240.}
\label{fig05}
\end{figure*}

To obtain the radio spectral index $\alpha$ between two frequencies, in each point over the area of the source, we need values of fluxes at the same coordinates $(\alpha,\delta)$. As we have used data with different resolutions, first we had to reduce them to the same resolution, and then to apply Eq. (\ref{equ02}). For that purpose we used bilinear interpolation for resampling observations with higher resolution (608 MHz) to the nearest existing coordinates at lower resolution (326 MHz). In that way, the data were comparable, and we were able to calculate spectral index.

The spectral index is used for classification of radio sources and studying the origin of radio emission:

\begin{itemize}[label=-,nosep]
\item If $\alpha > 0.1$ the emission is non-thermal (synchrotron) and it means that it does not depend on the temperature of the source, 
\item for $\alpha < 0$ it is thermal and depends only on the temperature of the source.
\end{itemize}

We calculated spectral indices between 326 and 608 MHz, over the whole area of DA 240, and we show how they change over this area in Fig. \ref{fig05}.

From the colorbar in Fig. \ref{fig05} we can read the values of radio spectral index $\alpha$, and we can notice that over the area of DA 240 it ranges from $\approx 0$ to positive values, meaning this: $\alpha > 0$ corresponds to non-thermal mechanism of radiation while $\alpha < 0$ corresponds to thermal mechanism of radiation. When the spectral index is zero, the flux density is independent of frequency, and the spectrum is said to be flat.

As is it can be seen from the presented radio-index map, the spectral index is almost always higher than zero, except in only few small parts where it is around zero. As expected, the largest values of spectral index are in the regions around hotspots (especially eastern one), while the lowest values are in vicinity of AGN, which is dominated by thermal radiation mechanism. This indicates that the non-thermal (synchrotron) emission is by far the most dominant radiation  mechanism over the whole source (except at AGN).

\section{Discussion and conclusions}

We used the available flux densities of DA 240 at 326 and 608 MHz to provide the spectral index distribution derived between these two frequencies. At both frequencies, the flux structure is characterized with obvious hotspots and the core. We can notice a large variations of flux density over the intensity map, with its highest value at the eastern component, which is dominant high brightness region. In the spectral index map this tendency is even more pronounced.

For the first time we give spectral index map of DA 240 between 326 and 608 MHz. We show that synchrotron radiation is the dominant emission mechanism over the majority of the area of the source, except in the central core.

Our investigation of the giant radio galaxy DA 240, i.e. of the flux density and spectral index distribution, is leading to the following conclusions:
\begin{itemize}
\item by using publicly available data (Leahy's atlas of double radio-sources, as well as the NED database), we were able to investigate flux densities at 326 MHz (92 cm) and 608 MHz (49 cm),
\item although we have developed method of calculation for main Galactic radio loops I-VI, it is applicable (and also rather efficient) to all SNRs, end to extragalactic radio sources, as well,
\item from our results, a remarkable feature can be noticed: the huge range of flux density over the intensity map, as well as the prominence of the eastern component,
\item there are two lobes, with a prominent hotspot in the north-east component, and a weaker one to the south-west,
\item the distribution of spectral index $\alpha$ enables to follow how $\alpha$ varies over the area, to read its values, and also to determine the origin of the radiation,
\item synchrotron radiation is the dominant emission mechanism over the whole area of the source.
\end{itemize}

\acknowledgements
This work is supported by Ministry of Science, Technological Development and Innovations of the Republic of Serbia through the Project contracts No. 451-03-47/2023-01/200002 and 451-03-47/2023-01/200017.


\end{document}